\documentclass[12pt,preprint]{aastex}

\lefthead{Devine \textit{et al.}}
\righthead{Supernova Remnants}

\begin{document}
\newcommand\HII{H\,{\sc ii}}
\newcommand\HI{H\,{\sc i}}
\newcommand\OI{[O\,{\sc i}] 63 $\mu$m}
\newcommand\CII{[C\,{\sc ii}] 158 $\mu$m}
\newcommand\CI{[C\,{\sc i}] 370 $\mu$m}        
\newcommand\SiII{[Si\,{\sc ii}] 35 $\mu$m}
\newcommand\Hi{H110$\alpha$}
\newcommand\He{He110$\alpha$}
\newcommand\Ca{C110$\alpha$}
\newcommand\kms{km~s$^{-1}$}
\newcommand\cmt{cm$^{-2}$}
\newcommand\cc{cm$^{-3}$}
\newcommand\Blos{$B_{los}$}
\newcommand\Bth{$B_{\theta}$}
\newcommand\Bm{$B_{m}$}
\newcommand\Bvm{$\mid\vec{B}\mid$}
\newcommand\BS{$B_S$}
\newcommand\Bscrit{$B_{S,crit}$}
\newcommand\Bw{$B_W$}
\newcommand\mum{$\mu$m}
\newcommand\muG{$\mu$G}
\newcommand\mjb{mJy~beam$^{-1}$}
\newcommand\jb{Jy~beam$^{-1}$}
\newcommand\dv{$\Delta v_{FWHM}$}
\newcommand\va{$v_A$}
\newcommand\Np{$N_p$}
\newcommand\np{$n_p$}
\newcommand\pp{$^{\prime\prime}$}
\newcommand\km{km~s$^{-1}$}
\newcommand\h{^{\rm h}}
\newcommand\m{^{\rm m}}
\newcommand\s{^{\rm s}}

\title{A Low Frequency Survey of the Galactic Plane Near
$\ell=11\arcdeg$: Discovery of Three New Supernova Remnants}

\author{C. L. Brogan\altaffilmark{1,2}, 
K. E. Devine\altaffilmark{3,4}, 
T. J. Lazio\altaffilmark{5},  N. E. Kassim\altaffilmark{5}, 
C.  R.  Tam\altaffilmark{6}, W.  F.  Brisken\altaffilmark{1}, 
K.  K. Dyer\altaffilmark{1,7}, and M.  S.  E.  Roberts\altaffilmark{6}}

\altaffiltext{1}{National Radio Astronomy Observatory, P. O. Box O, 1003 
Lopezville Road, Socorro, NM 87801; cbrogan@aoc.nrao.edu; 
wbrisken@aoc.nrao.edu; kdyer@aoc.nrao.edu}
\altaffiltext{2}{National Radio Astronomy Observatory Jansky Fellow}

\altaffiltext{3}{Carleton College, Department of Physics and Astronomy, 
Northfield, MN 55057}
\altaffiltext{4}{Current address: University of Wisconsin, Department of 
Astronomy, 5534 Sterling Hall, 475 North Charter Street, Madison, WI 53706; 
kdevine@astro.wisc.edu}

\altaffiltext{5}{Naval Research Laboratory, Remote Sensing Division, 
Code 7213, 4555 Overlook Avenue SW, Washington, DC 20375-5351; 
joseph.lazio@nrl.navy.mil; namir.kassim@nrl.navy.mil}

\altaffiltext{6}{Department of Physics, McGill University, 3600 University 
Street, Montreal QC H3A2T8, Canada; tamc@physics.mcgill.ca; 
roberts@physics.mcgill.ca}

\altaffiltext{7}{NSF Astronomy and Astrophysics Postdoctoral Fellow}

\begin{abstract} 

We have imaged a $\sim 1$~deg$^2$ field centered on the known Galactic
supernova remnant (SNR) G11.2$-$0.3 at 74, 330, and 1465 MHz with the
Very Large Array radio telescope (VLA) and 235 MHz with the Giant
Metrewave Radio Telescope (GMRT).  The 235, 330, and 1465 MHz data have a
resolution of $25\arcsec$, while the 74 MHz data have a resolution of
$\sim 100\arcsec$.  The addition of this low frequency data has allowed
us to confirm the previously reported low frequency turnover in the radio
continuum spectra of the two known SNRs in the field:  G11.2$-$0.3 and
G11.4$-$0.1 with unprecedented precision.  Such low frequency turnovers
are believed to arise from free-free absorption in ionized thermal gas
along the lines of site to the SNRs.  Our data suggest that the 74 MHz
optical depths of the absorbing gas is 0.56 and 1.1 for G11.2$-$0.3 and
G11.4$-$0.1, respectively.  In addition to adding much needed low
frequency integrated flux measurements for two known SNRs, we have also
detected three new SNRs:  G11.15$-$0.71, G11.03$-$0.05, and
G11.18+0.11.  These new SNRs have integrated spectral indices between
$-0.44$ and $-0.80$.  Because of confusion with thermal sources, the high
resolution (compared to previous Galactic radio frequency surveys) and
surface brightness sensitivity of our observations have been essential to
the identification of these new SNRs.  With this study we have more than
doubled the number of SNRs within just a $\sim 1$~deg$^2$ field of view
in the inner Galactic plane.  This result suggests that future low
frequency observations of the Galactic plane of similar quality may go
a long way toward alleviating the long recognized incompleteness of
Galactic SNR catalogs.

\end{abstract}


\section{INTRODUCTION}

Supernova explosions have a profound effect on the morphology, kinematics,
and ionization balance of galaxies, and possibly trigger new generations of
star formation.  However, based on statistical studies of the
incompleteness of Galactic SNR catalogs and predictions of the
Galactic SNR rate, there should be many more SNRs in our Galaxy \citep[$\sim
1000$;][]{Li1991, Tammann1994,  Case1998} than are currently known
\citep[$\sim 230$;][]{Green2002}.  This paucity is likely due in part to
selection effects acting against the discovery of the more mature, faint,
extended remnants, as well as, the very young, small remnants due to poor
sensitivity and spatial resolution at low frequencies where SNRs are
brightest \citep[c.f.][]{Green1991}.  These missing remnants are thought to
be concentrated toward the inner Galaxy where the diffuse nonthermal
Galactic plane emission coupled with the thermal emission from \HII\/
regions causes the most confusion.  A complete census of Galactic SNRs is
essential to understand the star formation history of our Galaxy.

Recent advances in both the instrumentation and software for low radio
frequency observations make sensitive, high resolution imaging in this
relatively unexplored waveband possible for the first time \citep[see e.g.]
[]{LaRosa2000}.  As a test case for a larger fully sampled mosaic of the
inner Galactic plane from $\ell=+4\arcdeg$ to $+20\arcdeg$ we imaged the
region around the known SNR G11.2$-$0.3 with the NRAO\footnote{The National
Radio Astronomy Observatory (NRAO) is a facility of the National Science
Foundation operated under a cooperative agreement by Associated
Universities, Inc.}  Very Large Array (VLA) at 74 and 330 MHz using data
from multiple configurations.  We have also obtained a VLA
mosaic at 1465 MHz of this region along with 235 MHz data from the Giant
Metrewave Radio Telescope (GMRT).  The final region of overlap of these
different data sets covers a $\sim 1$~deg$^2$ field of view, in which we
detect two known SNRs:  G11.2$-$0.3 and G11.4$-$-0.1 and identify three new
SNRs.  The results from these observations are presented below.

\section{OBSERVATIONS}

We have carried out observations of the Galactic plane centered on the
known SNR G11.2$-$0.3 at four frequencies:  1465, 330, 235, and 74 MHz.
The 1465, 330, and 74 MHz data were obtained with the VLA in multiple
configurations, while the 235 MHz data were obtained from the GMRT.  The
FWHP primary beams of the single pointing 330, 235, and 74 MHz data are
$2.5\arcdeg$, $1.5\arcdeg$ and $11.5\arcdeg$, respectively.  Data at all
four frequencies were acquired in spectral line mode in order to aid in
radio frequency interference (RFI) excision and minimize bandwidth
smearing.  The parameters for these observations are listed in Table 1.
The data from all four frequencies were reduced and imaged using the
wide-field imaging capabilities within the NRAO AIPS (Astronomical Image
Processing Software) package.  More information on wide-field imaging
techniques can be found in \citet{Cornwell1992}.  Some additional details
of wide-field Galactic plane imaging can be found in \citet{LaRosa2000}.

The initial antenna based gain and phase corrections for the 235, 330, and
1465 MHz data were calculated in the usual way from observations of the
phase calibrator 1833$-$210 and either 3C286 or 3C48 for flux calibration.
Unfortunately, there are no point like calibrators at 74 MHz that are
strong enough to dominate the total flux in the large 74 MHz primary beam
($11.5\arcdeg$).  For this reason, the flux scale and a single instrumental
phase at 74 MHz were estimated from observations of Cygnus~A using a
publicly available model of its source structure.  Since Cygnus~A is far
from the target field, initial estimates for the phase as a function of
time, as well as positional information for the target field were obtained
by using the VLA 330 MHz image as the model in the first round of 74 MHz
phase only self-calibration \citep[see][for further details specific to 74
MHz VLA data reduction]{Kassim2003}.  At all four frequencies, following
the initial calibrat-based corrections, several iterations of
self-calibration were applied to data from each configuration or
observation listed in Table 1 in order to improve the dynamic range.  The
calibrated visibility data from all of the available observations for each
frequency were then concatenated into a single $U$-$V$ data set, after
which one further amplitude and phase calibration was performed.

The 1465 MHz VLA image is composed of a three pointing mosaic, with the
pointings arranged in a triangle with G11.2$-$0.3 near its center.  The
FWHP primary beam of each pointing is $0.5\arcdeg$.  The three pointings
were calibrated and imaged separately and then combined with the linear
mosaicing task FLATN within AIPS.  During the FLATN process, the
individual pointings were corrected for the response of the primary beam
and then masked at the 25\% power point.  The final 1465 MHz image covers
a $\sim 1$~deg$^2$ area.  After correction for the primary beam response,
the 330, 235, and 74 MHz images were subimaged to the same field of view.

To correct for the gain compression that the GMRT suffers when observing
toward bright sources (i.e.  the inner Galactic plane), an additional gain
correction factor must be applied to the 235 MHz data after imaging.  This
factor can be estimated at the time of the observations by comparing the
total power data for the calibrators with and without the automatic level
attenuators on for each antenna.  At the time of the second GMRT 235 MHz
observation (see Table 1), this empirical factor had an average value of 4
(this factor is not available for the first GMRT observation).  However,
the integrated flux density measurements of the known SNRs (G11.2$-$0.3 and
G11.4$-$0.1) after multiplying the final image by this correction factor
were excessively high given their known spectral indices.  This discrepancy
may be due in part to the amplitude self-calibrations that were applied to
the 235 MHz data before the gain correction factor, or it is possible that
the correction factor for the first observing run may be different from
that of the second.  In order to better calibrate the 235 MHz gain, we have
used the spectral index between 330 and 1465 MHz at each pixel for
G11.2$-$0.3 and G11.4$-$0.1 to create an image of the expected flux density
at 235 MHz for each pixel.  This approach is justified since the continuum
spectra of SNRs above 150 MHz are well characterized by a constant
power-law spectrum \citep[see e.g.][]{Kassim1989a}.  Comparison of the
expected 235 MHz image with the observed (uncorrected) image yielded an
average correction factor of 2.7.  The final 235 MHz image was multiplied
by 2.7, before primary beam correction.  The corrected 235 MHz image was
then convolved to a resolution of $25\arcsec$.

For the 1465 and 330 MHz VLA data, care was taken to match the $U$-$V$
coverage so that each composite data set had an intrinsic resolution
of approximately $25\arcsec$.  The images were then convolved to
exactly $25\arcsec$, but the convolution was minimal.  The 1465 and
330 MHz data are not sensitive to structures larger than $\sim
15\arcmin$. The VLA 74 MHz data have a resolution of $145\arcsec\times
92\arcsec$.  The 74 MHz image does not include A-configuration data
because exceptionally poor ionospheric conditions rendered them
uncalibratable.  Since the 74 MHz resolution only marginally resolves
the SNRs, and as described above, the amplitude calibration of the 235
MHz data is a bit uncertain, data from these two frequencies were only
used for integrated flux measurements.

\section{RESULTS} 

Figures 1a and 1b show our VLA 1465 MHz and 330 MHz images with
$25\arcsec$ resolution of a $\sim 1$~deg$^2$ area centered at
G11.2$-$0.3.  The rms noise levels in these two images are $\sim 1$
\mjb\/ at 1465 MHz and $\sim 3$ \mjb\/ at 330 MHz, while the dynamic
range in these two images are 470 and 340, respectively.  The known SNRs
G11.2$-$0.3 and G11.4$-$0.1 are labeled on Figure  1a for reference along
with the locations of the three new SNR candidates designated A, B, and
C.  The Galactic coordinates for the new SNR candidates are A:
G11.15$-$0.71, B:  G11.03$-$0.05, and C:  G11.18+0.11.  These positions
have been taken from the approximate center of the SNR shells, and for
the purpose of this paper two significant digits (instead of the usual
one) are used to help distinguish the new candidate SNRs from the two
known remnants.  Given the strong morphological and spectral evidence
presented below, we will henceforth refer to these sources as ``new
SNRs'', rather than SNR candidates.

Integrated flux density measurements were made for the four frequencies
in our study for all five SNRs.  Since the images have different
noise levels, and several of the sources have regions of ``breakout''
morphology that tend to gradually fade into the noise (i.e.  not
sharp shells) it was difficult to determine where the boundaries of the
SNRs should be drawn.  Therefore, in order to get comparable flux density
estimates, the $4\sigma$ (12 \mjb\/) contour of the 330 MHz image (see
Figure  1b) was used to define the SNRs for all of the integrated flux
measurements.  The errors on these flux density measurements were
estimated by ${\rm (\#~independent~beams)^{1/2}}\times 3\sigma$ at each
frequency.  Upper limits for nondetections were estimated by ${\rm
(\#~independent~beams)^{1/2}}\times 5\sigma$.  The rms noise of the 235
MHz and 74 MHz data are 8 \mjb\/ and 150 \mjb\/, respectively.  When
necessary, an extra contribution from the background was subtracted from
the integrated flux density.  This contribution was estimated by
integrating the flux in a nearby source free region with the same
approximate area as the source.  The background level correction was only
necessary for the 74 and 235 MHz images.  In addition, the error
estimates for the 235 MHz flux densities have had an additional $5\%$
error added to account for the extra uncertainty in the 235 MHz gain
calibration described above.

Flux densities from the literature have also been included in our analysis
for all cases in which an error estimate is available and the error is
less than 20\%.  Figures 2 and 3 show the integrated radio continuum
spectra for the two known SNRs in our field, G11.2$-$0.3 and G11.4$-$0.1.
Figures 4-6 show the integrated radio continuum spectra for the new SNRs
A (G11.15$-$0.71), B (G11.03$-$0.05), and C (G11.18+0.11). 
Further details for each SNR are described below in \S 3.1.

For each SNR, the integrated spectral index was calculated using a
weighted least squares fit using the data points shown in Figures 2-6
(except for the 74 MHz upper limits).  Due to the method described
above to calibrate the 235 MHz gain, the data points at 235 MHz were
not used to fit the spectral index for G11.2$-$0.3 and G11.4$-$0.1
(i.e.  since these 235 MHz flux densities were used to set the 235 MHz
gain scale).  The data for G11.2$-$0.3 and G11.4$-$0.1 were fit to the
equation
\begin{equation}
S_{\nu}= S_{1000}\left(\frac{\nu} {1000~{\rm MHz}}\right)^{\alpha}
~{\exp}\left[-\tau_{1000} \left(\frac{\nu} 
{1000~{\rm MHz}}\right)^{-2.1}\right],
\end{equation}
where $S_{1000}$ and $\tau_{1000}$ are the flux density and optical depth
at a fiducial frequency of 1000 MHz, respectively, and $\alpha$ is the
integrated spectral index \citep{Dulk1975, Kassim1989a}.  This equation
assumes a standard nonthermal constant power law spectrum and allows for a
thermal absorption turnover at lower frequencies \citep[see
e.g.][]{Dulk1975, Kassim1989a}.  The free-free continuum optical depth at
other frequencies can be estimated from $\tau_{\nu}=\tau_{1000}
\left(\frac{\nu} {1000~{\rm MHz}}\right)^{-2.1}$.  The new SNRs were deemed
to have too few data points to adequately constrain Eq.  1, instead they
were fit with a pure power law (i.e.  $S_{\nu}\propto \nu^{+\alpha}$).  The
solid lines in Figures 2-6 show the best fits for each SNR.

Figures 7a-e show spatially resolved spectral index maps toward
the five SNRs between 330 and 1465 MHz. Before calculating the
spectral index, the two images were masked at the $4\sigma$ level (12
and 4 \mjb\/, respectively). Due to the low level calibration and
positional errors that are inherent in these low frequency images,
small scale spectral index variations less than $\sim 0.1$ should be
viewed skeptically. However, these maps are quite useful for 
identifying regions of thermal (positive $\alpha$) vs. 
nonthermal emission (negative $\alpha$).

\subsection{Individual Sources}

\subsubsection{SNR G11.2$-$0.3}

The SNR G11.2$-$0.3 is thought to be coincident with the historical
supernova of A.D.  386 \citep{Clark1977}.  Although there is still some
ongoing debate over the reality of this association, it is clear that
G11.2$-$0.3 is quite young \citep{Roberts2003}.  The distance to this
remnant has been estimated from \HI\/ absorption measurements to be $\sim
5$ kpc \citep{Radhakrishnan1972, Green1988}.  In the radio, G11.2$-$0.3 has
a steep spectrum radio shell with a significantly flatter spectrum central
core \citep[see Figure 7a;][]{Tam2002, Morsi1987, Kothes2001}.  The soft
X-ray morphology of G11.2$-$0.3 follows that of the radio shell while there
is a nonthermal, hard X-ray core consistent with a pulsar wind nebula
\citep[PWN;]{ Vasisht1996}.  Subsequent X-ray observations have
confirmed the presence of a 65 ms X-ray pulsar near the center of the
remnant \citep{Torii1997, Kaspi2001, Roberts2003}.

\citet{Kothes2001} have studied the morphology of G11.2$-$0.3 between
10 and 32 GHz using the Effelsberg radio telescope and find that the data
are consistent with a radio shell spectral index of $\alpha_{S}\sim -0.57$
and a PWN spectral index of $\alpha_{P}\sim 0.0$, with the PWN contributing
about 1 Jy of flux to the integrated flux density of the remnant ($\sim
6\%$ at 1.4 GHz).  A higher resolution study of the spectral index variations
of G11.2$-$0.3 between 1465 and 4865 MHz using archival VLA data has been
carried out by \citet{Tam2002} in order to study the interaction of the
PWN with the shell.  These authors find that $\alpha_{S}\sim-0.56$ and that
the PWN region has $\alpha_{P}\sim -0.25$ and contributes about 1 Jy of
flux density at 1465 MHz.  However, this analysis may suffer from missing
short spacings, along with confusion from the SNR shell emission.

In order to account for the flatter spectrum of the core, 1 Jy has
been subtracted from the integrated flux measurements shown in Figure 2
for data with $\nu > 200$ MHz, assuming that $\alpha_{P}\sim 0.0$. We
have not subtracted 1 Jy below 200 MHz under the assumption that the
PWN could suffer from self-absorption at some frequency, as
well as the extrinsic free-free absorption observed for the SNR as a
whole. However, without more information about the spectral properties
of the PWN, this choice of cutoff frequency is somewhat
arbitrary. It is clear from the fit shown in Figure 2 that there is
still a small excess of flux at higher frequencies, suggesting that
$\alpha_{P}$ is actually somewhat inverted (i.e. $\alpha_{P}> 0.0$). A
more detailed study of the spectral characteristics of the PWN will be
presented in Tam et al., in prep. based on the new VLA 330 MHz image
and 1465 MHz mosaiced image presented here, along with new 4835 and 8600
MHz mosaics of this region.

The integrated continuum spectrum shown in Figure 2 for G11.2$-$0.3 shows a
sharp down turn for $\nu<200$ MHz.  This spectral behavior was also noted
by \citet{Kassim1989a} in a survey to determine the spectral properties of
SNR at low frequencies, and will be discussed further in \S 4.2.  The best
fit parameters to the data shown in Figure 2 (after subtracting 1 Jy as
described above) are $\alpha=-0.570\pm 0.002$ for the shell and an optical
depth at 74 MHz of $\tau_{74}\sim 0.56$. The spatially resolved spectral 
index of G11.2$-$0.3 is shown in Figure 7a. The flatter spectrum of the 
PWN is clearly visible in this image.

The nature of the weak, extended ($\sim 15\arcmin$) linear feature that lies
tangent to the southern edge of G11.2$-$0.3 (see Figures 1a, b, and 7a) is
unkown.  The radio spectrum of this extended emission is fairly steep with
spectral indicies of $\sim -0.5$.  Moreover, there are no extended infrared
sources near this linear structure (see \S 4.1).  Together, these facts
suggest that the emission is nonthermal in origin.

\subsubsection{SNR G11.4$-$0.1}

G11.4$-$0.1 was first identified as an SNR by \citet{Caswell1975}.  The
distance to this SNR is currently unknown.  The best fit parameters for the
integrated continuum spectrum shown in Figure 3 are $\alpha=-0.59\pm 0.01$
and an optical depth at 74 MHz of $\tau_{74}\sim 1.1$.  The spatially
resolved spectral index map for G11.4$-$0.1 shown in Figure 7b suggests the
presence of a number of flatter ($\alpha\sim -0.4$) than average
($\alpha\sim -0.6$) spectral index filaments running through the SNR.
Similar spectral flattening toward filamentary structures has recently been
observed toward a number of mature SNRs including W28 and W49B
\citep{Dubner2000, Moffett1994}.  Such flattening may be indicative of
first order Fermi shock acceleration at the sites of strong shocks
\citep[see e.g.][]{Jun1999}.

\citet{Dubner1993} suggested that this SNR could be composed of two
superposed remnants due to its morphology in $63\arcsec\times 44\arcsec$
resolution 1465 GHz VLA images.  From the 330/1465 MHz spectral index map
shown in Figure 7b, it seems likely that G11.4$-$0.1 is composed of a
single nonthermal structure.  The point source near RA
$=18{\rm^h}10{\rm^m}36{\rm^s}$, DEC $=-19\arcdeg08\arcmin00\arcsec$ (J2000)
in Figs 1a, b and 7b has a flat radio spectrum ($\alpha\sim +0.1$), and
since it does not have an infrared counterpart, it is unlikely to be
thermal in origin (see \S 4.1) and is most likely an unrelated
extragalactic source.  It could also be a pulsar, but since this region of
the plane was observed in the Parkes Multibeam pulsar survey
\citep{Morris2002} it seems unlikely that a pulsar with such a strong radio
continuum would have gone undetected.  While strong to be a heretofore
undetected pulsar, the point source only contributes a few tens of mJy to
the integrated SNR flux density so no effort was made to subtract it from
the integrated flux measurements.

\subsubsection{New SNR A: G11.15$-$0.71}

The integrated continuum spectrum for the new SNR A located at
G11.15$-$0.71 is shown in Figure 4, while a map of its spatially resolved
spectral index between 330 MHz and 1465 MHz is shown in Figure 7c.  At 330
and 1465 MHz G11.15$-$0.71 appears as a rim brightened partial shell on its
northern end.  G11.15$-$0.71 has also been weakly detected in the 2695 MHz
Bonn survey \citep[][this data point is included on Figure 4]{Reich2001}
but has not previously been spatially resolved or recognized as a
nonthermal source.  Its integrated spectral index is quite steep with
$\alpha= -0.82\pm 0.01$.

\subsubsection{New SNR B: G11.03$-$0.05}

New SNR B at G11.03$-$0.05 appears as a complete shell at 330 MHz and a
partial shell at 1465 MHz (see Figures 1a, b).  A plot of its integrated
continuum spectrum is shown in Figure 5, while the morphology of its
spatially resolved 330/1465 MHz spectral index is shown in Figure 7d.
G11.03$-$0.05 has an integrated spectral index of $-0.60\pm 0.01$.  The
double source at RA $=18{\rm^h}09{\rm^m}43\fs2$, DEC
$=-19\arcdeg26\arcmin28\arcsec$ (J2000) (just west of the SNR shell
structure) with a positive spectral index has also been detected in a 3cm
radio recombination line (RRL) survey by \citet{Lockman1989} with the NRAO 140
ft telescope with a $3\arcmin$ beam.  This RRL has a velocity of 18.5
\kms\/ and $\Delta v\sim 24$ \kms\/.  Other tracers of star formation have
also been detected in this region with similar velocities including a
molecular cloud in CS and methanol masers \citep{Bronfman1996, Walsh1998}.
Confusion from the associated \HII\/ region likely explains why this SNR
has not been discovered in lower resolution surveys.

\subsubsection{New SNR C:  G11.18+0.11}

Figure 6 shows the integrated continuum spectrum for SNR C at G11.18+0.11,
while Figure 7e shows the morphology of the spatially resolved 330/1465 MHz
spectral index.  The integrated spectral index for this SNR candidate is
$-0.44\pm 0.01$.  The region of positive spectral index near the northern
end of candidate C (J2000 RA$=18{\rm^h}09{\rm^m}54{\rm^s}$, DEC
$=-19\arcdeg10\arcmin12\arcsec$) is coincident with a RRL detected by
\citet{Lockman1989} with a velocity of 7.3 \kms\/ and $\Delta v\sim 54$
\kms\/ using the NRAO 140 ft at 3 cm (to within their $3\arcmin$ beam).  It
is difficult to estimate the contribution of this thermal component to the
integrated SNR flux due to the absence of spatially resolved data at a
higher frequency.  For example, an unresolved source is apparent at the
position of the SNR candidate in the 2695 MHz Bonn survey images but it is
confused with both the \HII\/ region traced by the RRL, as well as, the
steep spectrum point source just east of the SNR shell that is likely
extragalactic (see Figure 7e).  For this reason, no effort was made to
remove this thermal component from the integrated flux densities, and the
reported $\alpha$ should be viewed as a lower limit.

\section{DISCUSSION} 

\subsection{The G11.2$-$0.3 Field at Other Wavelengths}

Comparisons of high resolution radio and infrared Galactic plane images are
quite useful in distinguishing different components of the ISM \citep[see for
example][]{Cohen2001}.  Figure 8 shows a $\sim 20\arcsec$ resolution
mid-infrared 8.28 \mum\/ image from the {\em Midcourse Space Experiment
Satellite} (MSX) Galactic Plane Survey of the G11.2$-$0.3 field with the 4
\mjb\/ 1465 MHz contour superposed \citep[see][for details about the MSX
image processing]{Price2001}.  In the Galactic plane, the 8.28 \mum\/
emission is dominated by contributions from PAH molecular line emission, and
thermal emission from warm dust around evolved stars, and regions of star
formation (normal, and ultracompact \HII\/ regions).  The details of the
origin of the diffuse mid-IR emission is less certain but may be associated
with the diffuse warm ionized medium (WIM).  The 1465 MHz radio continuum
image has contributions from both nonthermal, as well as, ionized thermal
sources.  In Figure 8 sources with both strong infrared and radio detections
are likely to be normal or ultracompact \HII\/ regions; unresolved sources
with radio but no infrared emission are nonthermal and are most likely
extragalactic but could also be very young SNRs, pulsars, or unresolved PWNe;
unresolved sources with infrared, but no radio continuum emission are most
likely normal and evolved stars; extended infrared sources without radio
continuum emission trace the very youngest stages of star formation where the
surrounding medium has not yet been ionized (regions that are optically thick
even at 8.28 \mum\/ can appear as mid-IR voids with no radio emission, e.g.
in the SW corner of Figure 8); and finally, extended radio continuum emission
without strong infrared emission must be nonthermal and may be SNRs
\citep[see][for a more detailed discussion of these classifications and some
caveats]{Cohen2001}.  There are a few cases where extended infrared
emission has been detected toward SNRs, but this only seems to occur when an
SNR shocks a nearby molecular cloud \citep[i.e.  3C~391,][]{Reach2002}.

A low ratio of infrared to radio continuum flux has been used for some time
as a tool to identify SNR candidates in Galactic plane surveys
\citep{Whiteoak1996}.  However, at the comparatively poorer resolutions used
in past surveys ($>1\arcmin$) application of this test could be ambiguous in
confused regions of both thermal and nonthermal emission.  The high
resolution images shown in Figure 8 dramatically demonstrate the location of
the Galactic plane running through the eastern half of the image, as well as,
the confusion associated with star formation in this region, particularly
near the new SNRs B:  G11.03$-$0.05 and C:  G11.18+0.11.  Many of the sources
in our 330/1465 MHz spectral index maps (Figures 7a-e) with $\alpha>0.1$ have
counterparts at 8.28 \mum\/ and are likely to be \HII\/ regions.  The radio
point sources located toward the SW corner of G11.4$-$0.1 and to the east of
SNR C (see \S 3.1.2 and \S 3.1.5) are not coincident with infrared emission
and therefore, are statistically likely to be extragalactic although as
mentioned above they could also be pulsars.  No extended infrared features
(above the diffuse Galactic infrared background) are detected toward the five
SNRs discussed here.  There is also no evidence for an infrared counterpart
to the extended linear radio continuum feature that lies tangent to the
southern edge of G11.2$-$0.3 (\S 3.1.1).

The ASCA and ROSAT X-ray archives were also searched for possible
correlations with the SNRs in the G11.2$-$0.3 field.  With the exception of
G11.2$-$0.3 itself, whose X-ray properties have been discussed by a number
of authors \citep[see e.g.][]{Vasisht1996}, none of the other SNRs in the
field were detected by either ASCA or ROSAT.  However, there is a region of
diffuse X-ray emission in between SNRs B:  G11.03$-$0.05 and C:
G11.18+0.11 in an image from the ASCA Galactic plane survey (47 ks exposure
with the GIS instrument).  An ASCA X-ray image spanning the energy range
2-10 keV in the region of SNRs B and C is shown in Figure 9.  The X-ray
data were processed using the technique described in \citet{Roberts2001}.
The origin of the diffuse X-ray emission near the center of Figure 9 is
uncertain, however, the radio pulsar PSR J1809$-$1917 described in \S 4.3 is
coincident with a peak in the X-ray emission (see Figure 9).  Low signal to
noise diffuse emission is also coincident with the middle and northern
parts of G11.03$-$0.05.  The X-ray spectra derived from the ASCA data lack
sufficient counts to adequately fit their spectra.  It would be interesting
to investigate this region in the X-ray with higher sensitivity in order 
to determine the nature of the diffuse emission.

\subsection{Distances from the $\Sigma-D$ relation}

The $\Sigma-D$ relation (surface brightness vs.  diameter) has long been
used to assign distances to SNRs without alternative distance estimates,
based on the observed properties of SNRs at known distances.  Such
relations rely on the assumption that supernova explosion energies, ambient
densities, and the strength and evolutionary state of their magnetic fields
are similar.  It remains unclear to what extent these assumptions are true,
particularly in regard to the ambient density.  Indeed, while $\Sigma-D$
relations often provide a fairly tight correlation for a statistical
sample, their application to individual SNRs is fairly dubious
\citep[see][for further discussion]{Case1998}.
Despite this caveat, we have used the most recent determination of the
Galactic $\Sigma-D$ relation derived by \citet{Case1998}:
\begin{equation}
\Sigma_{1~GHz} = 2.07^{+3.10}_{-1.24} \times 10^{-17}D^{(-2.38\pm 0.26)},
\end{equation} 
to estimate the distances to the five SNRs discussed in this paper (see
Table 2).  The sample used to derive the fit in Eq.  2 includes data from
36 Galactic shell type SNRs.  \citet{Case1998} find that the
average fractional deviation of an individual SNRs distance in their
sample from that calculated from Eq.  2 is $\sim 40\%$.

Table 2 shows the parameters used to derive the SNR distances.  The
distance to G11.2$-$0.3 estimated from Eq.  2:  6 kpc, is fairly close to
that estimated from \HI\/ absorption (see \S 3.1.1).  Note that G11.2$-$0.3
was not included in the Case \& Bhattacharya derivation of the $\Sigma-D$
relation.  The estimated distances to the other SNRs are G11.4$-$0.1:  9
kpc, G11.15$-$0.71:  24 kpc, G11.03$-$0.05:  16 kpc, and G11.18+0.11:  17
kpc.  The range of these values assuming a $40\%$ fractional error are also
listed in Table 2.  It is notable that the fraction of low surface
brightness compared to high surface brightness shell type remnants among
the known SNRs is much higher than the fraction used to derive Eq.  2
\citep[i.e.  few SNRs with low surface brightness have known
distances;][]{Case1998}.  Hence, Eq.  2 is significantly biased toward high
surface brightness shell type SNRs.  The effect of this bias on Eq.  2, and
hence our three new low surface brightness shell type remnants is
uncertain.  It is interesting that the high end of the distance ranges
presented in Table 2 (assuming a $40\%$ fractional error), would
unrealistically place all three of the new SNRs outside the bounds of the
Galaxy, perhaps suggesting that the distance to these SNRs are not well fit
by Eq.  2.  Also, note that the cases of G11.15$-$0.71 and G11.18+0.11 are
particularly uncertain (independent of the accuracy of Eq.  2) since the
angular size ($\theta$) of these two SNRs are difficult to estimate since
they only appear as partial shells in the current images.  This analysis
suggests that the three new SNRs may lie on the far side of the Galaxy, but
\HI\/ absorption measurements are needed to better constrain their
distances.

\subsection{Pulsar Association?}

In addition to the pulsar associated with G11.2$-$0.3 described in \S
3.1.1, one other pulsar has been discovered within the confines our low
frequency field of view.  PSR J1809$-$1917 was discovered in the Parkes
Multibeam Pulsar Survey by \citet{Morris2002} and is located at
RA$=18{\rm^h}09{\rm^m}43\fs2$, DEC $=-19\arcdeg17\arcmin38\arcsec$ (J2000).
The position of the pulsar is indicated on Figure 9.  It has a pulse period
of 83 ms, a flux density at 1400 MHz of 2.5 mJy, dispersion measure DM= 197
pc cm$^{-3}$, a dispersion measure distance of $\sim 4$ kpc, and a
characteristic age of 50 kyr.  We do not detect the pulsar at either 330 or
1465 MHz.  This is not surprising for the 1465 MHz data because the noise
rises to $\sim 2$ \mjb\/ near the edge of the image (due to the primary
beam correction).  However, if we assume a typical pulsar spectral index of
$-2$ the predicted 330 MHz flux density is 45 \mjb\/ which should be easily
detectable in our 330 MHz image (rms noise 3 \mjb\/).  Alternatively, the
pulsar's spectral index could be lower than average ($\alpha=-1$ predicts
$S_{330}= 11$ \mjb\/).  In the current ATNF pulsar
catalog\footnote{http://www.atnf.csiro.au/research/pulsar/psrcat/}, $13\%$
of of the 283 pulsars with a measured spectral index have $\alpha>-1$.

The characteristic age estimate (50 kyr) assumes that the birth spin period
is negligible compared to the current spin period.  However, recent studies
on a small number of pulsars, for which an independent age estimate can be
made, suggest that the birth period is actually between 20 to 130 ms
\citep[see e.g.][]{Kaspi2001, Kramer2003}.  Thus for young pulsars that
currently have short spin periods (i.e.  little spin down has yet
occurred), a fundamental assumption of the characteristic age calculation
is violated.  This implies that the true ages of these pulsars could be
significantly younger than suggested by the characteristic age.  For
example, if the birth period of PSR J1809$-$1917 was $\sim 50$ ms, and we
assume that the spin down is dominated by dipole radiation with a braking
index of 3, then the true age of the pulsar is $\sim 25$ kyr, half the
characteristic age \citep[see][]{Kramer2003}.

PSR J1809$-$1917 is located about $8.5\arcmin$ from the center of G11.03$-$0.05
(new SNR B) and $6\arcmin$ from the center of G11.18+0.11 (new SNR C).  If
these SNRs are located $\sim 17$ kpc away (see Table 2) compared to the
$\sim 4$ kpc distance estimated for the pulsar, then obviously there can be
no association.  However, as described in \S 4.2 the uncertainty in
distances derived from the $\Sigma-D$ relation are considerable.  Also,
given that the pulsar is fairly young and the notorious difficulty with
establishing SNR/pulsar associations, it is worthwhile to determine whether
the pulsar could have reasonably moved to its current location from the
center of either remnant (assuming they are at the same distance).  No
independent distance has yet been obtained for PSR J1809$-$1917, so we will
assume the dispersion measure pulsar distance of 4 kpc.  This distance is
based on a model for the electron distribution along the line of sight to
the pulsar and is probably good to within $\sim 25\%$ 
\citep{Taylor1993}.  Assuming that the pulsar is 50 kyr old, the pulsar's
transverse speed would need to be $\sim 200$ \kms\/ to originate from
G11.03$-$0.05 and $\sim 140$ \kms\/ for G11.18+0.11.  Such transverse
velocities are quite reasonable for pulsars, and would still be plausible
if the true age of the pulsar is younger than the characteristic age as
described above, or the pulsar is more distant.  For example, from analysis
of the velocity distributions of a large sample of pulsars,
\citet{Arzoumanian2002} predict that $\sim 50\%$ of
pulsars have velocities $\gtrsim 500$ \kms\/.  Further assessment of the
reality of a possible association with either remnant will have to await
more reliable distance estimates for the SNRs.

\subsection{Low Frequency Turnovers for G11.2$-$0.3 and G11.4$-$0.1}

The integrated continuum spectra shown in Figures 2 and 3 for G11.2$-$0.3
and G11.4$-$0.1 show clear signs of a low frequency turnover.  Their
optical depths at 74 MHz of 0.56 and 1.1, respectively, are higher than
those found by \citet{Kassim1989a} by about $40\%$.  Because our new
spectra include critical new data points at 74 and 330 MHz and we have
chosen to only use integrated flux densities from the literature with error
estimates that are less than $20\%$, our estimated optical depths are more
accurate.  The three new SNRs also show indications of a low frequency
turnover based on their 74 MHz upper limits, which we hope to confirm with
future higher resolution 74 MHz data.  From a survey of the integrated
continuum spectra of 47 Galactic SNRs, \citet{Kassim1989a} found that $\sim
2/3$ of the observed remnants show a low frequency turnover below $\lesssim
200$ MHz.  This turnover is thought to be due to free-free absorption by
intermediate temperature ($\sim 3,000-8,000$ K) and density ($\sim 1-10$
\cc\/) ionized thermal gas along the line of sight to the SNRs \citep[see
e.g.][and references therein]{Kassim1989a}.  Based on the fact that such
absorption is not seen toward all SNRs, and does not seem to depend on
distance, it is unlikely that the absorbing medium is a distributed
component of the ISM, but is more likely associated with discrete sources.
Kassim estimates that the filling factor of ionized material with the
required electron densities and temperatures is $\lesssim 1\%$.

A number of origins for the absorbing medium have been suggested.  Two
obvious sources of Galactic ionized gas:  the warm ionized medium (WIM) and
\HII\/ regions do not provide the appropriate physical conditions.  The WIM
has electron densities that are too low by a factor of $\sim 10$ to 100 and
its filling factor is too high \citep[$> 10\%$;][]{Kulkarni1987}.  Normal
\HII\/ regions have electron densities and temperatures that are far too
high to account for the moderate low frequency optical depths that are
observed, and they would be easily detected in higher frequency continuum
observations.  One reasonable suggestion is that the extended envelopes of
normal \HII\/ regions (EHEs) might well provide the requisite temperatures
and densities \citep{Anantharamaiah1985,Anantharamaiah1986}.  It is unclear
whether such envelopes extend far enough into the ISM to account for a
large fraction of the SNRs showing low frequency absorption.

Alternatively, \citet{Heiles1996} suggest that Galactic ``worms'' may cause
absorption toward some sources.  Worms are thought to be created after the
first generations of an OB star cluster explode as supernovae, blowing a
large evacuated superbubble, the walls of which are then ionized by later
generations of OB stars.  The spatial extent of such worms can be several
degrees across.  \citet{Koo1992} have cataloged 118 Galactic worm
candidates from single dish \HI\/ and radio continuum data plus
correlations with IRAS infrared data.  All of the worm candidates lie
toward the inner Galaxy.  \citet{Heiles1996} find that in some cases worms
also emit RRLs and that the electron densities and temperatures expected
for worms are commiserate with those needed to produce free-free absorption
at low radio frequencies.

The low frequency absorption observed toward the SNR W49~B (G43.3$-$0.2) by
\citet{Lacey2001} could be a case of absorption by a worm.  An early
identification of the existence of ionized thermal gas toward W49~B came
from the detection of RRL lines at $\sim 60$ \kms\/ \citep{Pankonin1976}.
\citet{Lacey2001} subsequently found excellent agreement between the spatially
resolved morphology of the free-free absorption at 74 MHz and that of \HI\/
absorption at $\sim 60$ \kms\/.  There are no obvious \HII\/ regions near
W49B (in the plane of the sky) with a velocity of $\sim 60$ \kms\/ that
could be a candidate for producing an EHE.  Lacey et al.  suggest that the
absorption is due to a few compact \HII\/ regions (and their envelopes)
along the line of sight to W49~B itself, but it is difficult to see how the
cores of such \HII\/ regions could escape detection at higher frequencies
\citep{Moffett1994}.  Interestingly, one of the worms cataloged by
\citet{Koo1992} is in this general direction (GW 44.8$-$1.8) has an extent of
$\sim 2\arcdeg$ and velocities between 43 to 66 \kms\/.  Future RRL
observations may determine whether the physical conditions in the worm are
in agreement with those needed to produce the free-free absorption.

It also possible that some SNRs could create their own absorbing medium if
they happen to run into a nearby molecular cloud, assuming that the shock
is a fast, ionizing J-type shock.  Evidence for this case has recently been
discovered for the SNR 3C~391 which is known to be interacting with a
nearby molecular cloud \citep[see e.g.][]{Reach1999}.  The spatially
resolved morphology of the 74 MHz free-free absorption toward 3C~391
matches very well with that of mid-infrared fine structure atomic lines
that are known to be good tracers of ionic shocks \citep[Brogan et al., in
prep.;][]{Reach2002}.  Since all of these phenomena exist, it is likely
that each of them are responsible for at least some fraction of the
observed free-free absorption and only by studying a number of sources in
detail can the dominant process be determined.

The emission measures implied by our newly derived optical depths 
for G11.2$-$0.3 and G11.4$-$0.1 can be calculated from:
\begin{equation}
\tau(\nu)=1.643\times 10^5 a(T_e,\nu) \nu^{-2.1}~{\rm EM}~T_e^{-1.35},
\end{equation} 
where EM is the emission measure, $T_e$ is the electron temperature, and
$a(T_e,\nu)$ is the Gaunt factor and is $\sim 1$ at the temperatures and
densities discussed here \citep{Dulk1975}.  Assuming an electron
temperature of 5000 K, the average emission measure (EM) for absorbing gas
toward G11.2$-$0.3 and G11.4$-$0.1 at 74 MHz are $2.8\times 10^3$ cm$^{-6}$
pc and $5.6\times 10^3$ cm$^{-6}$ pc, respectively.  Accounting for the
lower $T_e$ used here, 5,000 K instead of 10,000 K, these emission measures
are similar to those found toward the SNR W49~B \citep{Lacey2001}.

There are no indications in the spectral index maps of G11.2$-$0.3 and
G11.4$-$0.1 between 330 and 1465 MHz of the morphology of the absorbing
medium.  This is not surprising given that the 330 MHz optical depths are
only 0.02 and 0.05 for G11.2$-$0.3 and G11.4$-$0.1, respectively.  There
is some indication in our $145\arcsec\times 92\arcsec$ resolution 74 MHz
image that G11.2$-$0.3 suffers from more absorption on its eastern side
than the western side, but this will need to be confirmed by higher
resolution 74 MHz data.  The \HII\/ region G11.11$-$0.4 located just
south of G11.2$-$0.3 (see Figure  1a), is a potential candidate to provide
an absorbing EHE if its distance along the line of site is closer to
us than that of the remnant.  The kinematics of the molecular gas
associated with this \HII\/ region are consistent with a location in the
3 kpc arm of the Galaxy, suggesting that the distance to the \HII\/
region is $\sim 5$ kpc, similar to that of the SNR 
\citep{Solomon1987, Kurtz1994}.  A detailed comparison of \HI\/
absorption toward both G11.2$-$0.3 and G11.11$-$0.4 will be needed to
better determine their relative distances. No known \HII\/ regions 
are in close enough proximity (in the plane of the sky) to G11.4$-$0.1 
(see Figure 1a) to provide an EHE candidate for this SNR.

Spatially resolved 74 MHz images have been crucial in determining the
nature of the absorbing medium for W49~B \citep{Lacey2001} and 3C~391
(Brogan et al.  in prep.).  Future spatially resolved 74 MHz images are
likely to be equally important in identifying the source of ionized 
thermal gas toward G11.2$-$0.3 and G11.4$-$0.1.

\subsection{Implications for the Number of Galactic SNRs}

A number of recent studies that range from analyzing the rate of
historical Galactic supernovae, the rate of supernovae in the Local
Group of galaxies, to the birth rate of OB stars within 1 kpc of the
sun suggest that the rate of Galactic supernovae is about 1 per 50
years \citep[see e.g.][]{Tammann1994}. If we assume that
supernova remnants can be distinguished from the Galactic background
nonthermal emission for about 50,000 to 100,000 yrs then there should
be on the order of 1000 to 2000 SNRs in our galaxy today. This
number is far below the number of currently known Galactic SNRs, 
$\sim 230$ \citep{Green2002}.

For some time, observational selection effects have been thought to be the
responsible for the low number of Galactic SNRs currently known \citep[see
e.g.][]{Green1991}.  These selection effects include (1) the difficulties
associated with detecting the more mature, large, faint remnants in the
inner Galaxy due to confusion with the Galactic synchrotron background
radiation; (2) the challenges involved with identifying small, young shell
type SNRs that can be easily confused with steep spectrum extragalactic
sources; (3) difficulties with distinguishing young and older shell type
remnants when they are confused with \HII\/ regions that lie along the line
of sight; and (4) about $5\%$ of SNRs known as plerions have no
identifiable shell, and have flat spectra so that they are difficult to
distinguish from \HII\/ regions, even in high resolution images \citep[see
e.g.][]{Helfand1989}.  The solution to the first problem is obviously to
make high surface brightness sensitivity observations, while the second
requires high spatial resolution in the hopes of resolving a candidate SNRs
shell morphology.  The solution to number (3) is to obtain equally
sensitive, high resolution multi-wavelength data so that sources of thermal
emission can be separated from nonthermal emission.  The solutions to
number (4) include looking for strong linear polarization, or an associated
pulsar.

In the $\sim 1$~deg$^2$ field centered on G11.2$-$0.3 we have overcome the
first three of the above mentioned problems by obtaining high surface
brightness sensitivity, high resolution, multi-wavelength data.  These data
have resulted in the discovery of three new SNRs.  All three new remnants
appear fairly old and at least based on the somewhat dubious $\Sigma-D$
relation, are located on the other side of the Galaxy (16 to 24 kpc).  We
have not identified any new young SNRs in this field.  However,
statistically this is not surprising since this field also contains the
historical remnant G11.2$-$0.3 from A.D.  386, and we have only sampled a
small fraction of the inner Galactic plane.  If we define ``young'' SNRs as
having ages less than 2,000 years, then assuming one SNR every 50 years
implies that there are $\sim 40$ such ``young'' remnants in the Galaxy.
There are only six known historical remnants (meaning that sightings of
their optical emission were recorded in historical documents):  the Crab,
SN1006, 3C~58, Tycho, Kepler, and G11.2$-$0.3, plus a handful of suspected
``young'' SNRs including CasA.  Thus there are on the order of 30 ``young''
SNRs that have yet to be discovered.  These SNRs are presumably missing
from the historical records because the optical emission from their initial
explosions were obscured by intervening material.  Indeed, five of the six
historical SNRs described above, are a degree or more off the plane.

Extrapolating from this approximately 1~deg$^2$ field of view, we can
estimate how many SNRs should be detected in our planned 330~MHz survey of
the inner Galaxy, from $+4\arcdeg \le \ell \le +20\arcdeg$ with $|b| <
1\arcdeg$.  For this estimate, we make the simplifying assumption that the
distribution of SNRs within this strip of the Galactic plane is fairly
uniform.  However, this assumption could be violated if some lines of site
pass through more spiral arms than others.  \citet{Cordes2003} have recently
completed a new analysis of the Galactic free electron distribution
determined from pulsar dispersion measurements \cite[this work extends and
improves upon the original study by][]{Taylor1993}.  From the Galactic
electron distribution and many other lines of evidence, these authors have
also developed a an updated model for the Galactic spiral arm structure.  The
\citet{Cordes2003} spiral arm model suggests that there are two potential
disparities in the total number of spiral arms in the longitude range
$+4\arcdeg \le \ell \le +20\arcdeg$.  These include the fact that (1) the
Norma arm does not appear to extend beyond $\ell>10\arcdeg$ on the near side
of the Galactic center; and (2) the Crux-Scutum Arm does not appear to extend
to longitudes smaller than $\ell=18\arcdeg$ on the far side of the Galactic
center.  An even greater discrepancy is possible if a line of site passes
through a tangent point of a spiral arm where gas, \HII\/ regions, and
presumably SNRs tend to pile up.  This may well be the true for lines of site
near $\ell=20\arcdeg$.  In any case, the region near $\ell=11.2\arcdeg$ does
not appear to be a particularly special one within the $+4\arcdeg \le \ell
\le +20\arcdeg$ range, suggesting that this field of view should not be
overdense with SNRs compared to the rest of the survey region.

The above caveats aside, the results from our $\sim 1$~deg$^2$ field toward
G11.2$-$0.3 suggest that near $b=0\arcdeg$, the density of SNRs in this
part of the Galactic plane could be as high as 5 per deg$^2$.  We will
sample approximately 32 deg$^2$ in our survey, but we assume that only the
$\sim 16$ deg$^2$ near $b=0\arcdeg$ (i.e.  $\mid b\mid\lesssim
0.5\arcdeg$) has an SNR density of 5 per deg$^2$, with the extra area
accounting somewhat for density variations.  These assumptions imply a
total SNR count in the $+4\arcdeg\lesssim\ell\lesssim+20\arcdeg$ and $\mid
b\mid\lesssim 1\arcdeg$ region of $\sim 80$.  This number is in reasonable
agreement with that predicted by \citet{Helfand1989} for this region of the
Galactic plane (63) based on a Galaxy wide treatment of SNR densities
\citep[also see] [and references therein] {Case1998}.  Within the bounds of
our future survey region, only 14 SNRs are currently known
\citep{Green2002}!  Only the actual results from our more extensive survey
will determine whether this estimate is realistic, but in any case it will
go along way toward determining the true distribution of Galactic SNRs.

\section{CONCLUSIONS}

In a $\sim 1$~deg$^2$ field of view centered on the known Galactic SNR
G11.2$-$0.3, we have identified three new SNRs.  Previously, this field was
thought to contain only two SNRs.  The integrated spectral indices of the
new SNRs:  A:  G11.15$-$0.71, B:  G11.03$-$0.05, and C:  G11.18+0.11 range
from $-0.80$ to $-0.44$.  Two of the new SNRs (G11.03$-$0.05 and
G11.18+0.11) have confusing \HII\/ regions nearby (along the line of sight)
that are not spatially resolved from the SNRs at the resolution of the Bonn
2695 MHz survey, and this fact is likely responsible for their not having
been previously identified.  We have also confirmed the low
frequency spectral turnover in the integrated radio continuum spectra of
G11.2$-$0.3 and G11.4$-$0.1 previously observed by \citet{Kassim1989a} that
is due to free-free absorption along the line of sight.  The addition of
our new data suggest that the optical depths toward these SNRs at 74 MHz
are 0.56 for G11.2$-$0.3 and 1.1 for G11.4$-$0.1.  The upper limits for the
74 MHz integrated flux densities for the three new SNRs also suggest that
these remnants suffer from absorption.  While quite uncertain, a recent
formulation of the $\Sigma-D$ relation predicts a nearly correct distance
to G11.2$-$0.3 (6 kpc) and distances of 9, 24, 16, and 17, kpc for
G11.4$-$0.1, G11.15$-$0.71, G11.03$-$0.05, and G11.18+0.11, respectively.

Based on this ``test field'' we can say with confidence that high surface
brightness sensitivity combined with high spatial resolution low
frequency radio observations are an effective tool in the quest to find
the ``missing'' Galactic SNRs -- at least the more evolved ones.  While
single dish surveys have been invaluable for finding the majority of SNRs
known today, they are ineffective at finding faint, older remnants due to
confusion with the Galactic background synchrotron emission, SNRs that
are confused with thermal sources, and young small SNRs due to
insufficient resolution.  Interferometers overcome these problems by
resolving out the Galactic background while providing high surface
brightness sensitivity (assuming the inclusion of short spacing data) and
high spatial resolution.  Indeed, with the VLA data presented here we
have more than doubled the number of SNRs in a $\sim 1$~deg$^2$ field of
view.

However, the addition of new SNRs in a single $\sim 1$~deg$^2$ field is
insufficient to determine whether the selection effects mentioned above
can, in fact, account for all the missing remnants, or whether there is a
significant missing link in our understanding of the SNR production rate.
In order to obtain a better statistical sample we are currently carrying
out a wider survey from $\ell=+4\arcdeg$ to $+20\arcdeg$ along the
Galactic plane at 330 MHz using the VLA in its B, C, and D
configurations.  The resulting images will have $\sim 5$ \mjb\/
sensitivity and $\sim 20\arcsec$ resolution.  In combination with the
planned extension of the FIRST survey at 1465 MHz to this region of the
Galactic plane (R.  Becker, private communication), we should be able to
go a long way towards answering this question.

\acknowledgments

We thank P.  Rao for giving us the 235 MHz GMRT data observed
on Sept.  23, 2001, and F.  Camilo for alerting us to the presence of the
pulsar PSR J1809$-$1917 within our field of view.  Basic research in radio
astronomy at the NRL is supported by the Office of Naval Research.  KKD
is supported by an NSF Astronomy and Astrophysics Postdoctoral Fellowship
under award AST-0103879.  This research has made use of the NASA/ IPAC
Infrared Science Archive, which is operated by the Jet Propulsion
Laboratory, California Institute of Technology, under contract with the
National Aeronautics and Space Administration.

% \section{Figure captions}

\begin{figure}[h!]
\epsscale{0.55}
\plotone{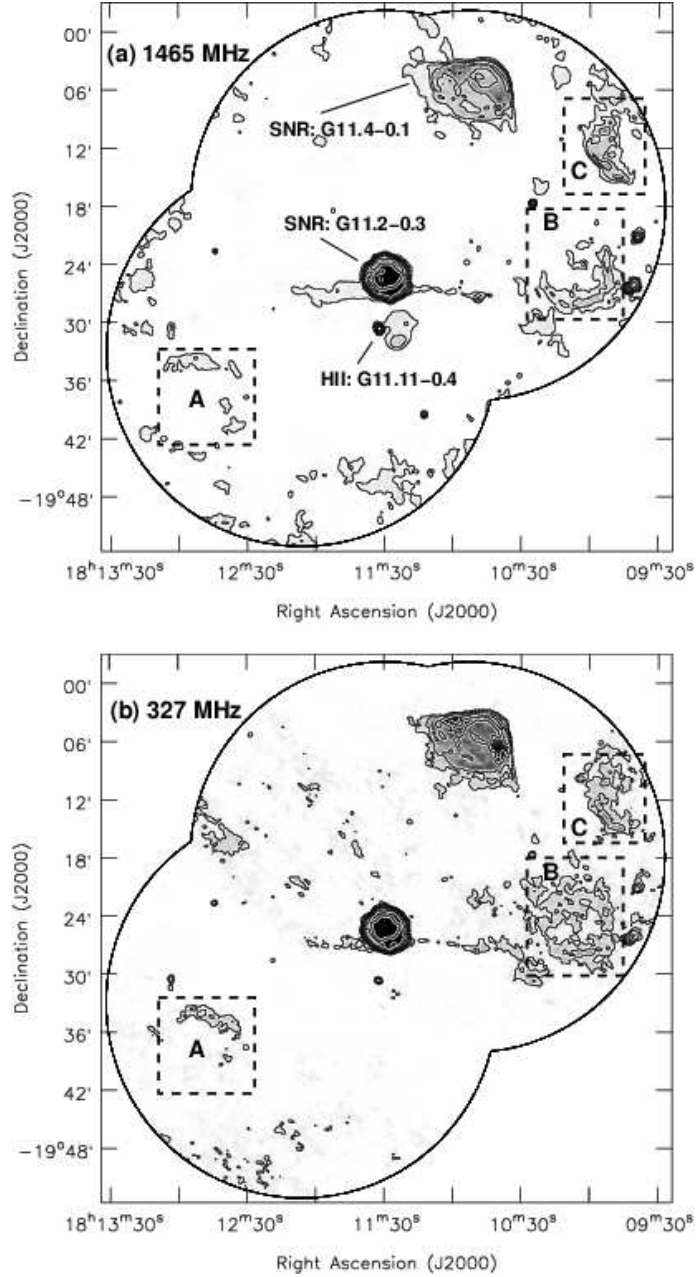}
\caption[]{(a) VLA 1465 MHz three pointing mosaic centered on the
known SNR G11.2$-$0.3 with $25\arcsec$ resolution.  The contour levels
are 4, 10, 20, 50, 100, 200, \& 400 \mjb\/.  The rms noise of this
image is $\sim 1$ \mjb\/ but varies somewhat as a function of position
due to the mosaicing procedure.  The sources identified with the
dotted boxes and labeled A, B, \& C are the three new SNR candidates.
(b) VLA image at 330 MHz centered on SNR G11.2$-$0.3 with $25\arcsec$
resolution.  The contour levels are 10, 20, 50, 100, 200, 400, \& 700
\mjb\/.  The rms noise of this image is $\sim 3$ \mjb\/.  The region
outside of the 1465 MHz mosaic has also been blanked on the 330 MHz
image for ease of comparison.}
\end{figure}

\begin{figure}[h!]
\epsscale{0.4}
\plotone{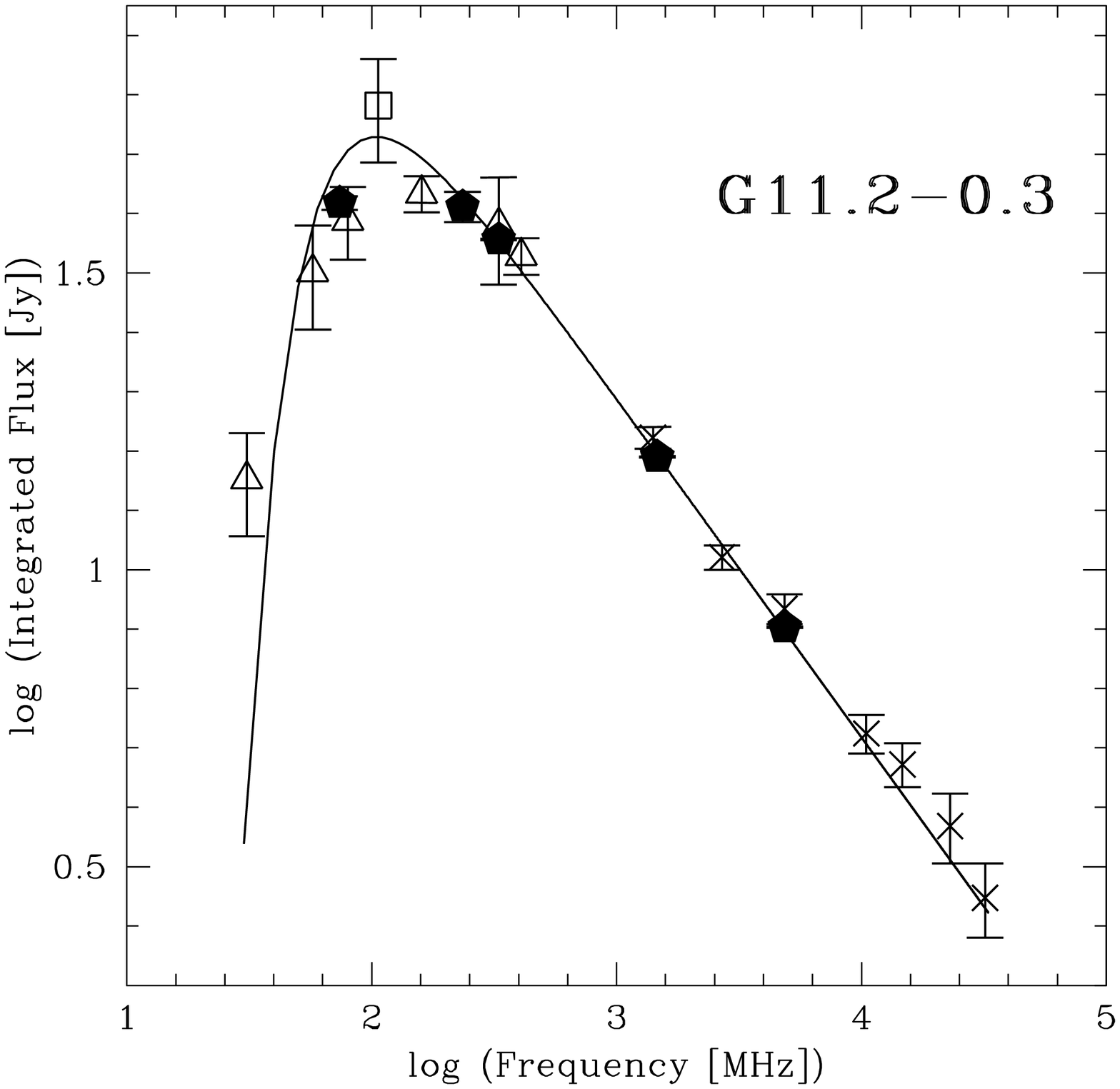}
\caption[]{Radio continuum spectrum for SNR G11.2$-$0.3. The 
best fit spectral index using Eq. 1 is $-0.570\pm 0.002$. 
The hexagon symbols are from the current work, the $\bigtriangleup$ symbols 
are from \citet{Kassim1989b, Kassim1992}, the single $\sq$ symbol is 
from \citet{Kovalenko1994}, and the the $\times$ symbols 
are from \citet{Kothes2001}. In
order to obtain the spectral index for only the shell of the SNR, 1 Jy 
has been subtracted from the integrated flux measurements for data with 
$\nu > 200$ MHz, assuming that $\alpha_{P}\sim 0.0$.}
\end{figure}

\begin{figure}[h!]
\epsscale{0.4}
\plotone{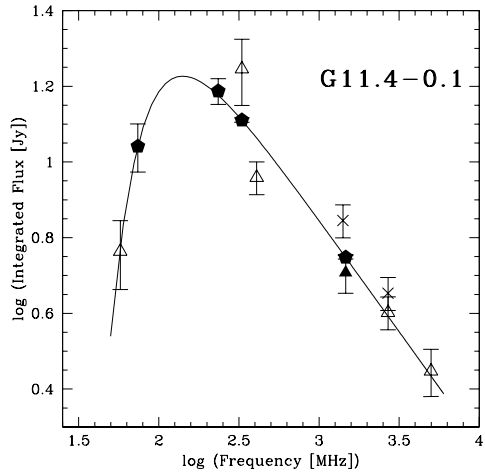}
\caption[]{Radio continuum spectrum for SNR G11.4$-$0.1.  The best fit
spectral index using Eq. 1 is $-0.59\pm 0.01$.  The hexagon symbols
are from the current work, the $\bigtriangleup$ symbols are from  
\citet{Kassim1989b, Kassim1992}, and the $\times$ symbols are from 
\citet{Reich1997, Reich2001}.  }
\end{figure}

\begin{figure}[h!]
\epsscale{0.4}
\plotone{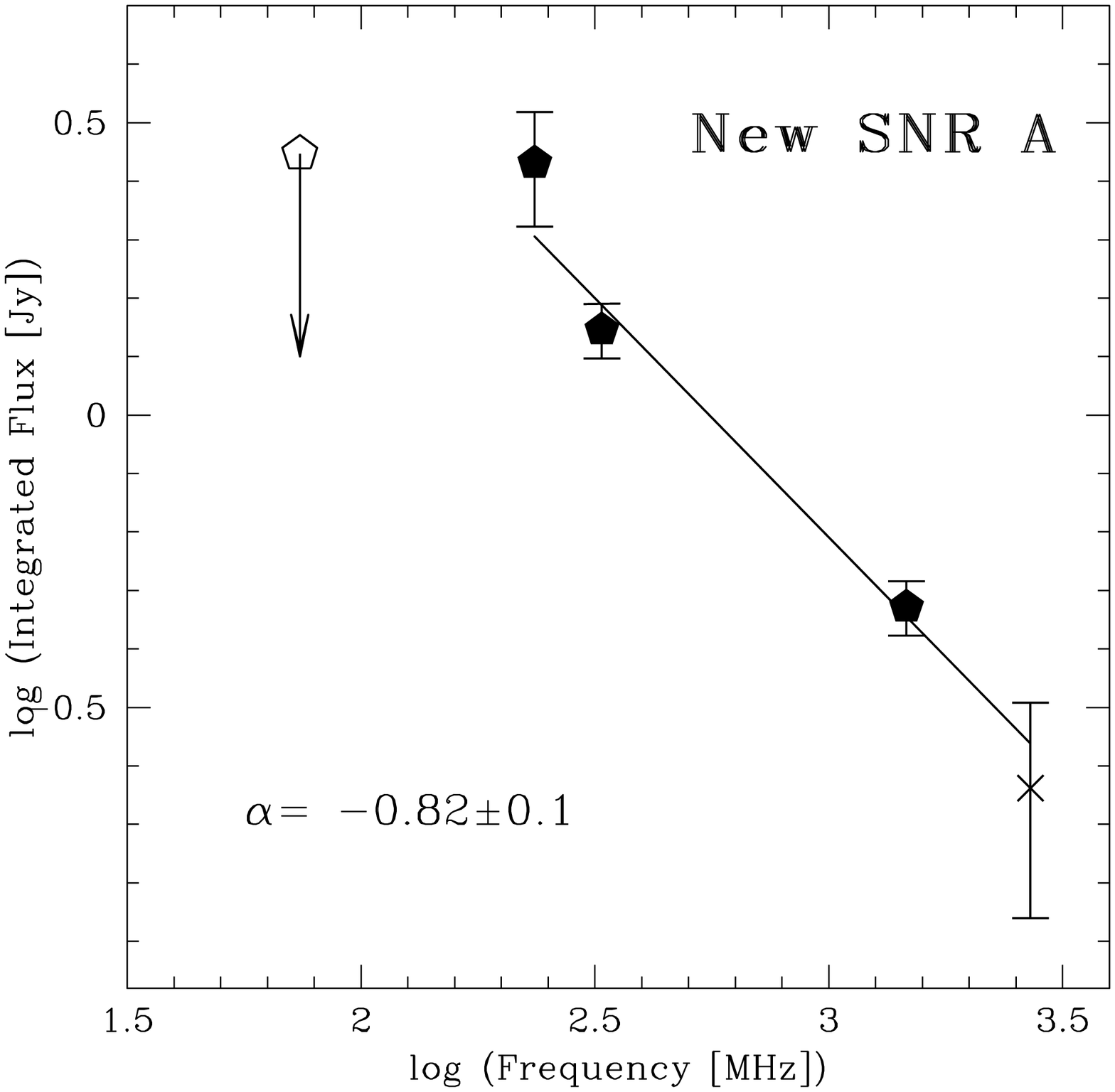}
\caption[]{Radio continuum spectrum for new SNR A: G11.15$-$0.71. The 
best fit power-law spectral index is $\alpha=-0.82\pm 0.01$. The hexagon symbols are from the 
current work while the integrated flux data point marked by the $\times$ 
symbol is from \citet{Reich2001}. The flux density marked by 
the open hexagon symbol at 74 MHz is an upper limit and was not used 
in the fit.}

\end{figure}
\begin{figure}[h!]
\epsscale{0.4}
\plotone{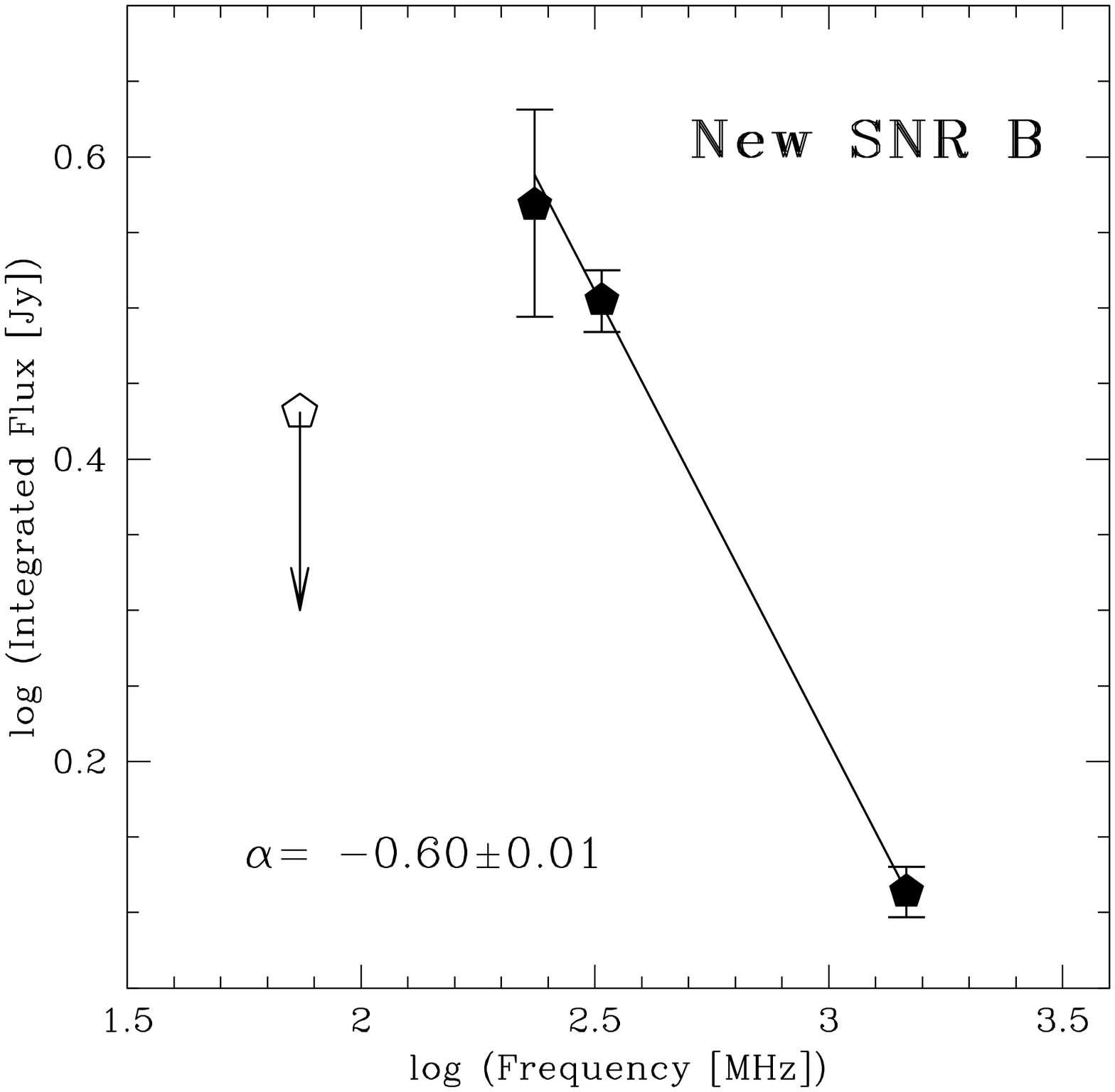}
\caption[]{Radio continuum spectrum for new SNR B: G11.03$-$0.05. The 
best fit power-law spectral index is $-0.60\pm 0.01$. The hexagon symbols are from the 
current work. The flux density marked by 
the open hexagon symbol at 74 MHz is an upper limit and was not used 
in the fit.}
\end{figure}

\begin{figure}[h!]
\epsscale{0.4}
\plotone{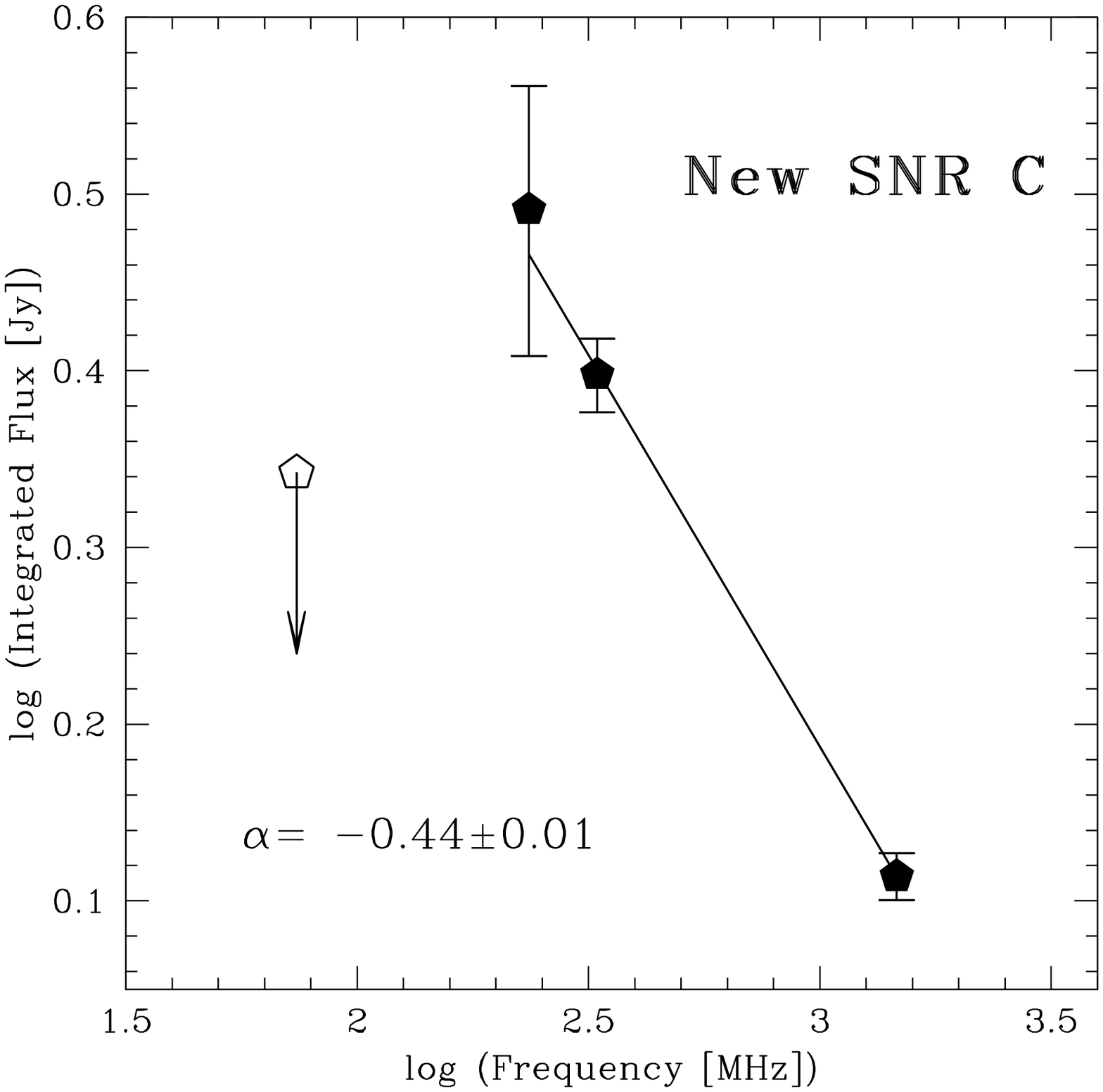}
\caption[]{Radio continuum spectrum for new SNR C: G11.18+0.11. The 
best fit power-law spectral index is $-0.44\pm 0.01$. The hexagon symbols are from 
the current work. The flux density marked by 
the open hexagon symbol at 74 MHz is an upper limit and was not used 
in the fit. }
\end{figure}

\begin{figure}[h!]
\epsscale{1.0}
\plotone{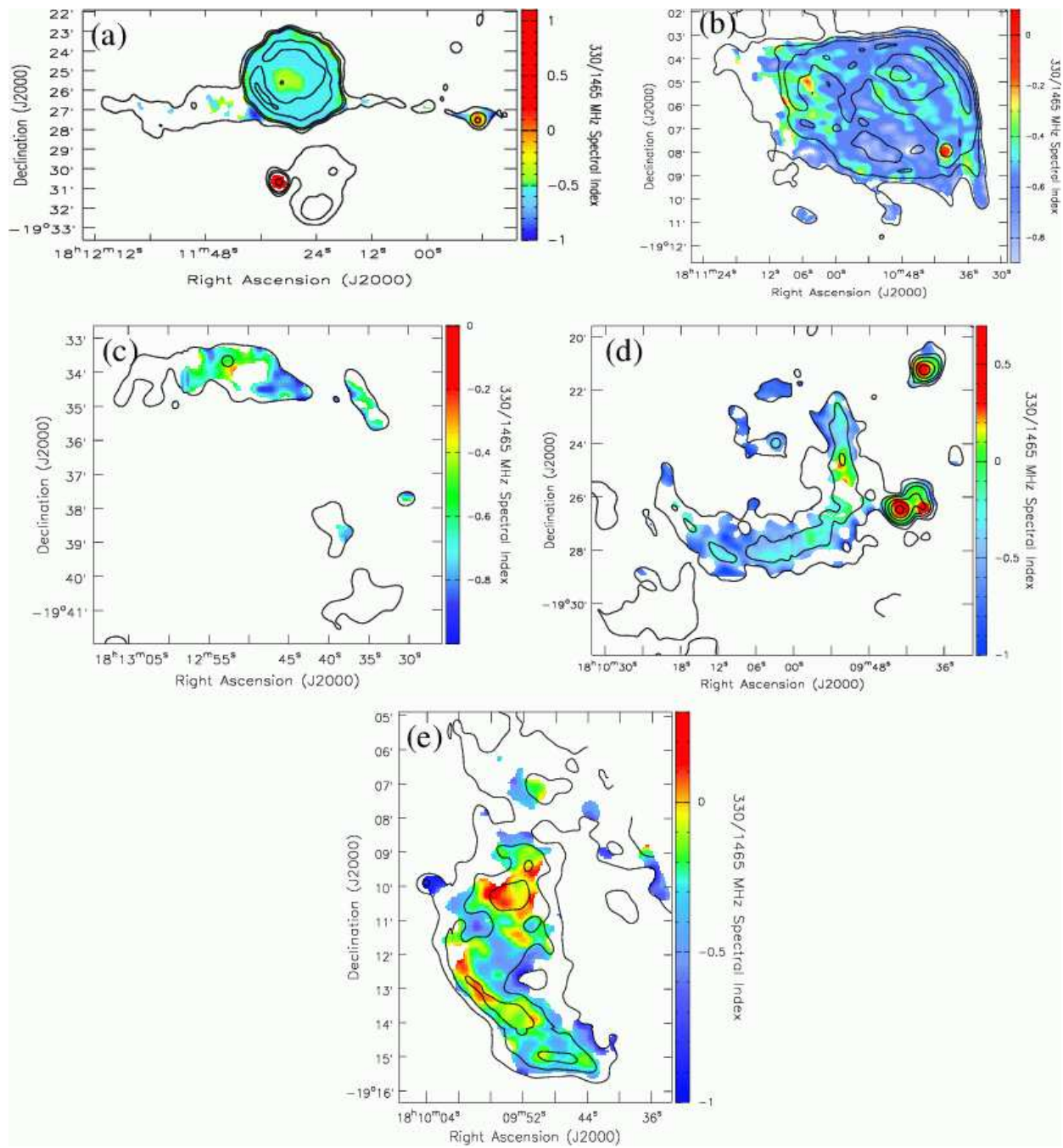}
\caption[]{Spectral index maps between 330 and 1465 MHz for (a) G11.2$-$0.3,
(b) G11.4$-$0.1, (c) new SNR A:  G11.15$-$0.71, (d) new SNR B:  G11.03$-$0.05,
and (e) new SNR C:  G11.18+0.11.  The 1465 MHz contours displayed in Figure 1a
are also shown.  The 330 and 1465 MHz continuum images were masked at 12 and 4
\mjb\/, respectively, before the spectral indices were calculated.  Note that
the color scale changes from image to image.}
\end{figure}

\begin{figure}[h!]
\epsscale{0.8}
\plotone{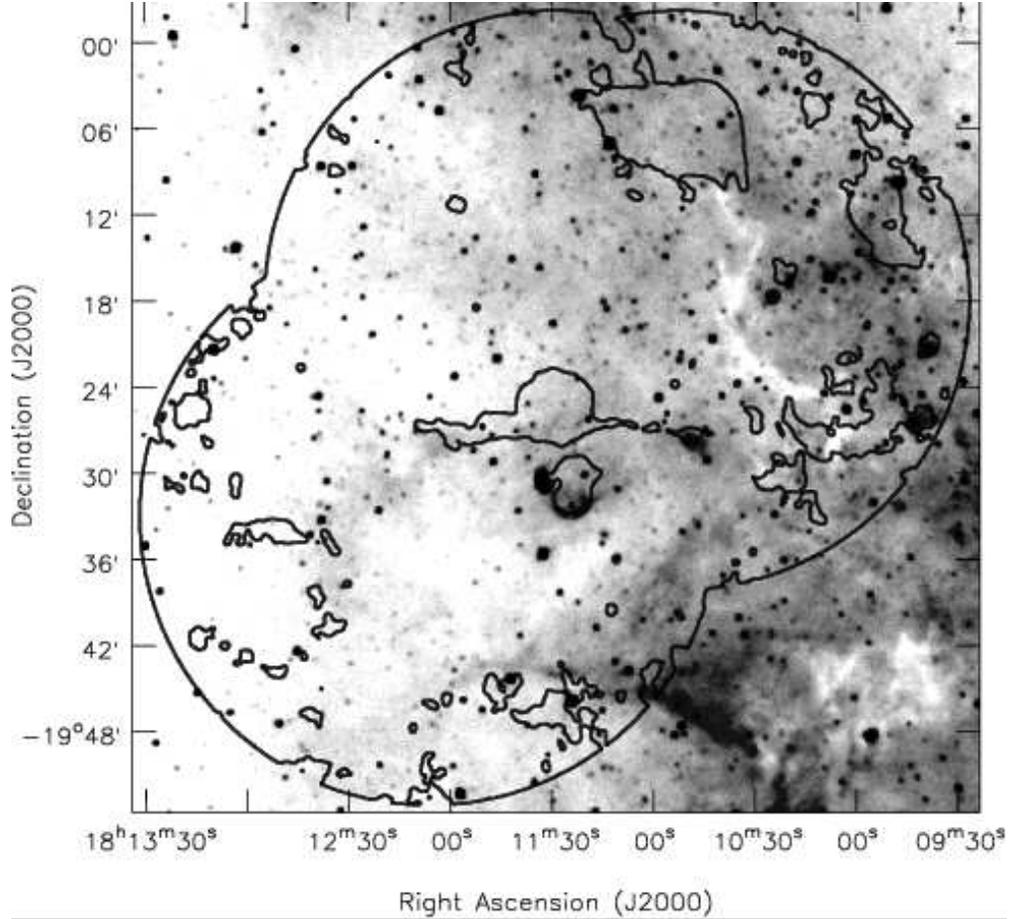}
\caption[]{MSX 8.28 \mum\/ image of the G11.2$-$0.3 field with the 4 
\mjb\/ 1465 MHz contour superposed. The resolution of the 8.28 \mum\/ image 
is $\sim 20\arcsec$. Note that many of the sources with 
positive spectral indices in Figure 7 are coincident with thermal 
gas traced by strong 8.28 \mum\/ emission.}
\end{figure}

\begin{figure}[h!]
\epsscale{0.5}
\plotone{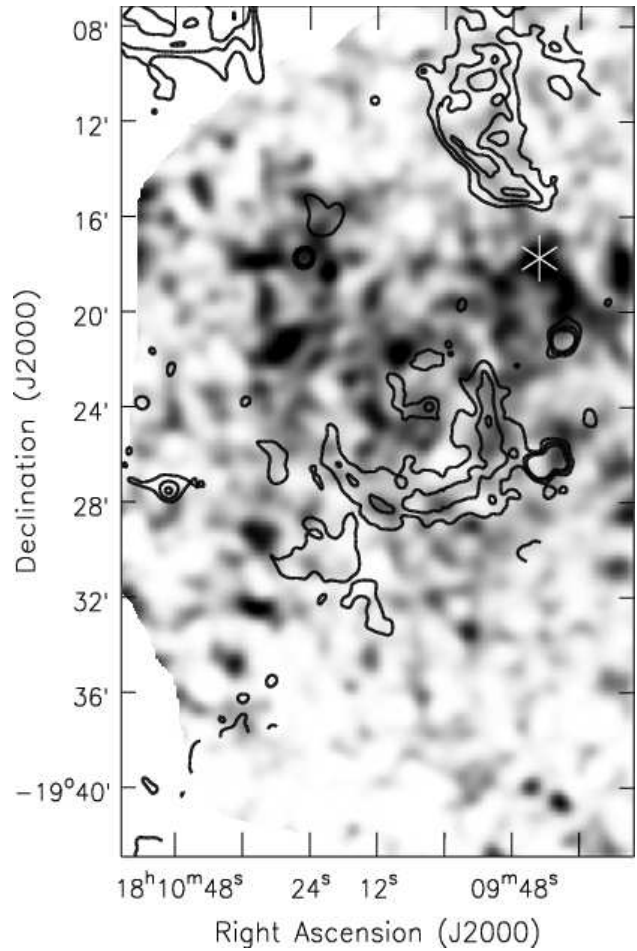} \caption[]{ASCA X-ray image in
the 2-10 keV range of the region containing the SNRs B:  G11.03$-$0.05 and
C:  G11.18+0.11 (see Figure 1a, b) with the 4, 10, and 20 \mjb\/ 1465 MHz
contours superposed.  The X-ray data has been convolved to the approximate
FWHM resolution of the GIS detector of $50\arcsec$.  The white * symbol
shows the location of pulsar PSR J1809$-$1917.  The image was processed as
described in Roberts et al.  2001.}

\end{figure}

\newpage

\begin{deluxetable}{lcccrc}
\small
\tablewidth{30pc}
\tablecaption{G11.2 Field Observational Parameters\label{tab1}}
\tablecolumns{5}
\tablehead{
\colhead{Date}  & \colhead{Instrument} & \colhead{Config.} &\colhead{Bandwidth} & \colhead{ Time~$^a$} 
\\ &   &  &\colhead{(MHz)} & \colhead{(Hours)}}
\startdata
\cutinhead{74/330 MHz parameters}
2001 Jan 13 & VLA & A &  --- /3 & 4.4\\
2001 Mar 1 & VLA & B & 1.5/3 & 1.7 \\
2002 Aug 20 & VLA & B & --- /12 & 1.5 \\
2001 Aug 28 & VLA & C & 1.5/3 & 0.75 \\
2001 Sep 28 & VLA & D & 1.5/3 & 1.7\\
\cutinhead{235 MHz parameters}
2001 Sep 23~$^b$ & GRMT & - & 16 & 1.5 \\
2002 May 05 & GMRT & - & 8 & 0.5\\
\cutinhead{1465 MHz parameters}
2001 Jun 27 & VLA & CnB & 25 & 0.3 \\
2001 Aug 03 & VLA & C & 25 & 0.3\\
2001 Sep 24 & VLA & DnC & 25 & 0.3\\

\enddata
\tablenotetext{a} {Approximate final time on source.}
\tablenotetext{b} {Data contributed by P. Rao}
\end{deluxetable}

\newpage

\begin{deluxetable}{lcccccc}
\scriptsize
\tablewidth{38pc}
\tablecaption{Derived SNR Parameters}
\tablecolumns{7}
\tablehead{
\colhead{Source Name} & \colhead{$\alpha$$^a$} & \colhead{$S_{1~{\rm GHz}}$} & \colhead{$\theta$} & 
\colhead{$\Sigma_{1~{\rm GHz}}$} &  \colhead{distance$^b$} &  \colhead{range$^c$}
\\ & & Jy & \colhead{(arcmin)} & \colhead{(W m$^{-2}$ Hz Sr$^{-1}$)} 
& \colhead{(kpc)} & \colhead{(kpc)}}
\startdata
G11.2$-$0.3 & $-0.57$ & 19.3$^d$ & 4.5 & $1.4\times 10^{-19}$ & 6 & 4$-$10\\
G11.4$-$0.1 & $-0.59$ & 7.0  & $7\times 9$ & $1.7\times 10^{-20}$ & 9 & 6$-$14\\
G11.15$-$0.71 (A) & $-0.82$ & 0.6 & $8\times 7$ & $1.7\times 10^{-21}$ & 24 & 17$-$40\\
G11.03$-$0.05 (B) & $-0.60$ & 1.6 & $6\times 7$ & $5.9\times 10^{-21}$ & 16 & 12$-$27\\
G11.18+0.11 (C) & $-0.44$  & 1.5 &$8\times 5$ & $5.8\times 10^{-21}$ & 17 & 12$-$28\\

\enddata
\tablenotetext{a} {Fitted spectral indices from integrated flux measurements 
shown in Figures 2-6.}
\tablenotetext{b} {Distances calculated using Equation 2.}
\tablenotetext{c} {Range of distances assuming a $40\%$ fractional error.}
\tablenotetext{d} {After subtracting 1 Jy for the PWN.}
\end{deluxetable}

\end{document}